\documentclass[aps,prl,twocolumn,showpacs,amsmath,floatfix]{revtex4}
\usepackage{graphicx}
\usepackage{dcolumn}
\usepackage{multirow}
\usepackage{amsmath}
\usepackage{subfigure}
\usepackage{braket}
\usepackage{color}
\usepackage{epstopdf}
\usepackage[normalem]{ulem}

\begin{document}


\title{Electronic origin of spin-phonon coupling effect in transition-metal perovskites}
\author{Hongwei Wang$^{1,2}$}
\author{Lixin He$^{1,*}$}
\author{Hong Jiang$^{3}$}
\author{Cameron Steele$^{2}$}
\author{Xifan Wu$^{2,*}$}
\affiliation{$^{1}$Key Laboratory of Quantum Information,
University of Science and Technology of China, Hefei, Anhui 230026, China }
\affiliation{$^{2}$Department of Physics, Temple University, 1925 N 12th Street, Philadelphia, Pennsylvania 19122, USA}
\affiliation{$^{3}$ Beijing National Laboratory for Molecular Sciences, College of Chemistry and Molecular Engineering,
Peking University, 100871 Beijing, China}

\date{\today}
\begin{abstract}
By applying Wannier-based extended Kugel-Khomskii model, we carry out first-principles calculations and
electronic structure analysis to understand the spin-phonon coupling effect in transition-metal perovskites.
We demonstrate the successful application of our approach to SrMnO$_3$ and BiFeO$_3$.
We show that both the electron orbitals under crystal field splitting and the electronic configuration
should be taken into account in order to understand the large variances of spin-phonon coupling effects
among various phonon modes as well as in different materials.
\end{abstract}


\maketitle


Spin-phonon coupling (SPC) is an important physical effect of multiferroic
materials~\cite{Spaldin2010}, in which the cross couplings between structural distortions
and magnetic orderings are closely associated with their key functionalities,
such as magnetoelectric coupling and magnetodielectric response.
Due to its fundamental and technological importance, the SPC effect
is currently under intense scientific investigations~\cite{FennieZn2006,FennieEu2006,Birol2013,Leenature2010}.

The mechanism in realizing SPC is not obviously
accessible since the structural distortion, in particular the development of ferroelectricity (FE),
does not necessarily induce a change of the magnetic interaction of the material.
SPC can be realized by the relativistic effect through the spin-orbital
interaction. FE can be induced by the spin spiral structure that breaks the inversion
symmetry~\cite{Tokura2014}, or the crystal structure in improper multiferroics can be
compatible with spin configurations generating the weak ferromagnetism (FM)~\cite{Wang2014,Das2014,Fennie2005}.
Unfortunately, the resulting electric and magnetic moments are
generally small. Recently, a new SPC mechanism has been discovered in
ABO$_3$ perovskites, in which A or B is magnetic transition metal
~\cite{Bhattacharjee2009,Lee2010,Lee2011,Kamba2014,Goian2015,Wang2012,Das2008, Birol2012}.
It was found that the low-lying phonon modes, particularly the polar ones,
are significantly softened when the spin coupling is changed from being
antiferromagnetic (AFM) to FM.

First-principles calculations have pioneered the search for SPC materials,
from which the differences of phonon frequencies between different spin configurations
can be predicted. As a result, a number of transition-metal perovskites with SPC effect
have been successfully identified~\cite{Bhattacharjee2009,Lee2010,Lee2011}. Notwithstanding the progress in the field,
several fundamental properties remain to be understood.
First, for a single material, the SPC strength varies significantly
among different phonon modes~\cite{LeeM2011,Hong2012,Garcia2011}. Second, SPC is not
observed as a general property for multiferroic materials. In particular, BiFeO$_3$ (BFO) as one
of the most studied room temperature multiferroic materials has surprisingly small SPC effect~\cite{Cheong2007,Kamba2007}.
Based on the Goodenough-Kanamori-Anderson (GKA) rules, the metal-oxygen-metal angles are 
often used to explain the SPC effect~\cite{Goodenough1955,Kanamori1959}; however, a phenomenological argument based
on the exchange angle only roughly captures the effect. Therefore, it is not either
accurate or conclusive. Precise assignments of electronic
processes involved in the magnetic exchange interactions and their couplings to
different phonon modes are keys in addressing the above questions.

In this letter, we elucidate the electronic origins of SPC effect by using SrMnO$_3$ (SMO) and BFO as examples.
In particular, we compute the superexchange (SE) interactions via the virtual electronic hopping
processes with a recently developed extended Kugel-Khomskii (KK) model~\cite{Kugel1982} based on maximally localized
Wannier functions (MLWFs)~\cite{Souza2001,Marzari2012}, in which the electronic screening is considered by constrained
random phase approximation~\cite{Vaugier2012}.
The SPC effect can be understood as the tendency towards the suppressed SE interaction
under the structural distortion along the phonon mode.
In these processes, the electronic structure plays a crucial role in the above.
On one hand, phonon modes that effectively change the hybridization between Mn-$3d$ and O-$2p$ represented by the
MLWFs are found to have strong SPC effect via the hopping integrals.
On the other hand, the rather different details in the virtual hopping processes
originating from the distinct electronic configurations in Mn$^{4+}$ and Fe$^{3+}$ ions explain
the much weaker SPC effect in BFO than that of SMO.
Our results  bridge the gap between the GKA phenomenological rule and electronic
structure of materials. Furthermore, it also provides important guidance to the search for new SPC materials.

\begin{figure*}[ht]
\includegraphics[height=3.6in]{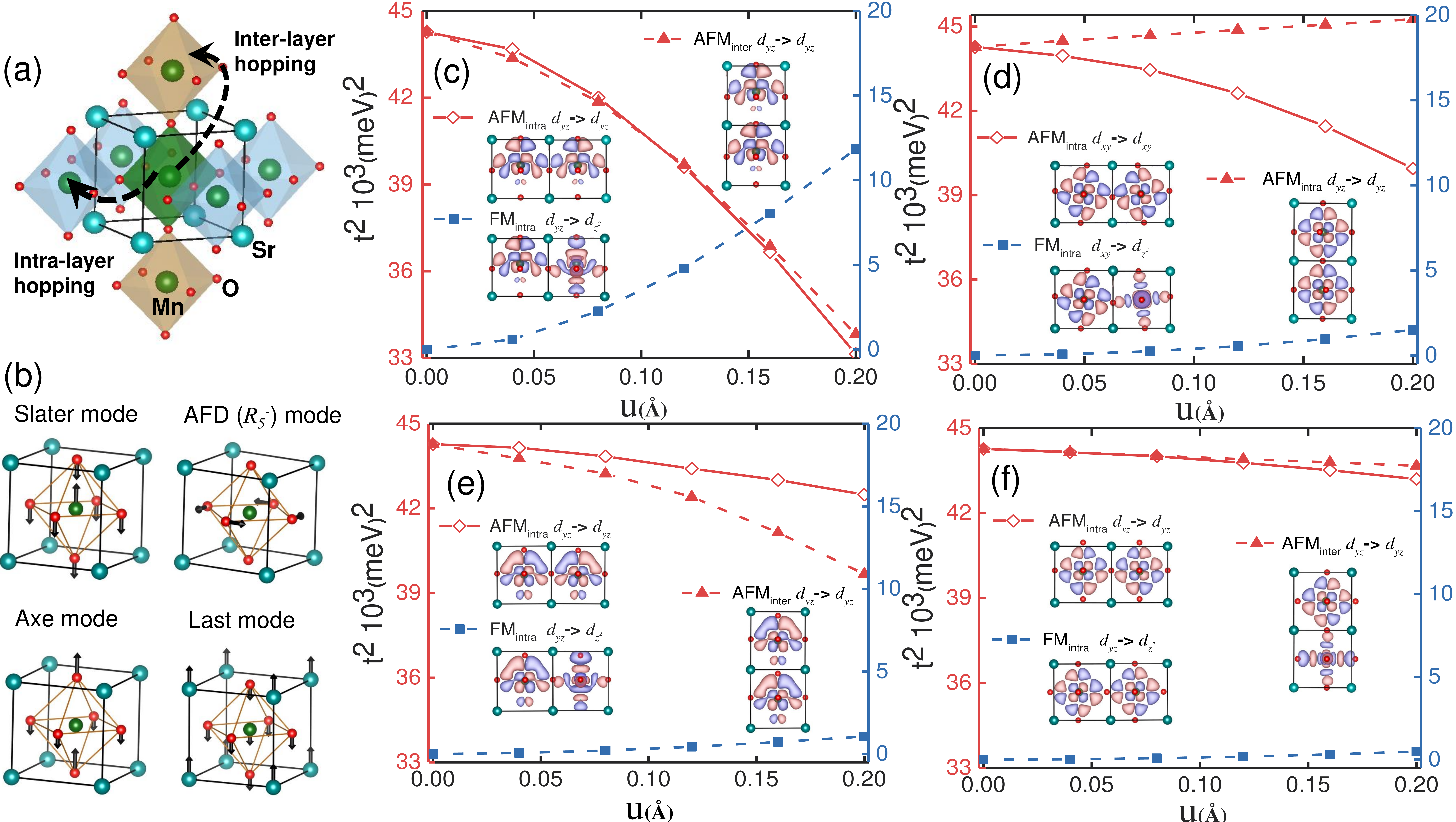}
\caption{\label{fig:superlattices} (color online)
(a) Schematic plots of the intra-layer and inter-layer virtual hopping processes involved in the
superexchange couplings in SMO. (b) Atomic displacements of the Slater, AFD, Axe, and Last phonon modes.
(c)-(f) Inter-layer and intra-layer hopping integrals involved in the AFM-type and FM-type superexchange couplings
as functions of  Slater, AFD, Axe, and Last phonon mode amplitudes in SrMnO$_3$ respectively. }
\end{figure*}

As a prototypical example, the SPC effect in SMO of cubic symmetry can be clearly seen by the softening
of low lying phonon frequencies when the spin coupling is changed from AFM to FM as shown in Table~\ref{Tab:response}.
In the above, the Slater, Axe, and Last modes~\cite{Slater1950,Last1957,Axe1967} are all polar modes originating from the Gamma point
instabilities; the antiferrodistortive (AFD) mode refers to the oxygen
octahedral rotation mode around [001] originating from the Brillouin zone boundary
instability~\cite{Woodward1997,Wang2016}.  The phonon frequency softening is a general trend for all the four modes in Table~\ref{Tab:response}.
However, the SPC strengths and therefore the magnitudes of the frequency shifts vary significantly.
As a result, $\Delta \omega_{\rm Slater} >\Delta \omega_{\rm AFD} >\Delta \omega_{\rm Axe}>\Delta \omega_{\rm Last}$.

\begin{table*}[ht]
\begin{center}
 \caption{Computed phonon frequencies $\omega$ and their shifts $\delta \omega$ between G-AFM and FM spin orderings
for Slater, AFD, Axe, and Last modes of SrMnO$_3$. For one  Mn atom,
$J^{''}_{\parallel,(\alpha, \beta)}$ ($J^{''}_{\parallel}$) and $J^{''}_{\perp,(\alpha, \beta)}$ ($J^{''}_{\perp}$)
represent the quadratic coefficients (meV/$\AA^2$)  of individual (total) intra-layer and inter-layer superexchange couplings
as functions of frozen mode amplitudes for the above four modes.
$J^{''}_{\rm T}=2 \times J^{''}_{\parallel}+J^{''}_{\perp}$ denotes the sum of the quadratic coefficients.
Values in parentheses were from Ref.~\cite{Hong2012}.}
\begin{tabular}{c|rccccc|ccccc|ccccc|ccccc}
  \hline
   \hline
      &    & \multicolumn{5}{c|}{Slater Mode} & \multicolumn{5}{c|}{AFD Mode}
& \multicolumn{5}{c|}{Axe Mode} & \multicolumn{5}{c}{Last Mode}
 \\
  \hline
 \multirow{2}{*}{$\omega$}   &    & \multicolumn{2}{c}{G-AFM } & \multicolumn{2}{c}{FM } & \multicolumn{1}{c|}{$\Delta \omega$ }
&\multicolumn{2}{c}{G-AFM } & \multicolumn{2}{c}{FM } & \multicolumn{1}{c|}{$\Delta \omega$ }
&\multicolumn{2}{c}{G-AFM } & \multicolumn{2}{c}{FM } & \multicolumn{1}{c|}{$\Delta \omega$ }
&\multicolumn{2}{c}{G-AFM } & \multicolumn{2}{c}{FM } & \multicolumn{1}{c}{$\Delta \omega$ }
  \\
   &    & \multicolumn{2}{c}{231(217) } & \multicolumn{2}{c}{103$i$(112$i$) } & \multicolumn{1}{c|}{-334 }
&\multicolumn{2}{c}{70$i$(71$i$)} & \multicolumn{2}{c}{112$i$(127$i$) } & \multicolumn{1}{c|}{-42 }
&\multicolumn{2}{c}{525(516) } & \multicolumn{2}{c}{518(504) } & \multicolumn{1}{c|}{-7 }
&\multicolumn{2}{c}{172(171) } & \multicolumn{2}{c}{171(170) } & \multicolumn{1}{c}{-1 }
  \\
     \hline
   &    & \multicolumn{3}{c}{\scriptsize{AFM type} } & \multicolumn{2}{c|}{\scriptsize{FM type} }
& \multicolumn{3}{c}{\scriptsize{AFM type} } & \multicolumn{2}{c|}{\scriptsize{FM type} }
& \multicolumn{3}{c}{\scriptsize{AFM type} } & \multicolumn{2}{c|}{\scriptsize{FM type} }
& \multicolumn{3}{c}{\scriptsize{AFM type} } & \multicolumn{2}{c}{\scriptsize{FM type} }
  \\
       &    & $d_{xy}$ &  $d_{xz}$ & $d_{yz}$ & $d_{x^{2}-y^{2}}$ & $d_{z^{2}}$
       & $d_{xy}$ &  $d_{xz}$ & $d_{yz}$ & $d_{x^{2}-y^{2}}$ & $d_{z^{2}}$
       & $d_{xy}$ &  $d_{xz}$ & $d_{yz}$ & $d_{x^{2}-y^{2}}$ & $d_{z^{2}}$
       & $d_{xy}$ &  $d_{xz}$ & $d_{yz}$ & $d_{x^{2}-y^{2}}$ & $d_{z^{2}}$
 \\
  \hline 
  \multirow{3}{*}{{$J^{''}_{\parallel,(\alpha, \beta)}$ }}
  & $d_{xy}$    & -8.5 & 0 & 0 & 0 & 0   & -40.1 & 0 & 0 & 8.6 & 1.9   & 0.4 & 0 & 0 & 0 & 0 & 1.3 & 0 & 0 & 0 & 0   \\
   & $d_{xz}$   & 0 & -39.3 & 0 & 23.2 & 14.7  & 0 & -1.5 & 0 & 0 &0 &  0 & -6.1 & 0 & 0.2 & 1.3 & 0 & -3.3 & 0 & 1.0 & 0.6 \\
 & $d_{yz}$    & 0 & 0 & -39.3 & 23.2 & 14.7 & 0 & 0 & -1.5 & 0 & 0 &  0  & 0  & -6.1 & 0.2  & 1.3 & 0 & 0 & -3.3 & 1.0 & 0.6 \\
  \hline
\multirow{3}{*}{{$J^{''}_{\perp,(\alpha, \beta)}$}}
 &$d_{xy}$               & 0  & 0 & 0  & 0  & 0  & 0  & 0 & 0  & 0  & 0  & 0  & 0 & 0  & 0  & 0 & 0  & 0 & 0  & 0  & 0 \\
   & $d_{xz}$   & 0 & -71.6 & 0 & 0 & 0   & 0 & 3.1 & 0 & 0 & 0   & 0 & -31.0 & 0 & 0 & 0  & 0 & -4.1 & 0 & 0 & 0\\
 & $d_{yz}$    & 0 & 0 & -71.6 & 0 & 0  & 0 & 0 & 3.1 & 0 & 0    & 0 & 0 & -31.0 & 0 & 0  & 0 & 0 & -4.1 & 0 & 0\\
  \hline
 $J^{''}_{\parallel}$  & \multicolumn{6}{c|}{$-$362.1 } &\multicolumn{5}{c|}{$-$107.1} &\multicolumn{5}{c|}{$-$29.9} &\multicolumn{5}{c}{$-$16.6}\\
 $J^{''}_{\perp}$  & \multicolumn{6}{c|}{$-$143.2 } &\multicolumn{5}{c|}{6.1} &\multicolumn{5}{c|}{$-$62.0} &\multicolumn{5}{c}{$-$4.9}\\
 $J^{''}_{\rm T}$  & \multicolumn{6}{c|}{$-$469.3  } &\multicolumn{5}{c|}{$-$101.0 } &\multicolumn{5}{c|}{$-$91.9 } &\multicolumn{5}{c}{ $-$21.4}\\
   \hline
   \hline
\end{tabular}
\label{Tab:response}
\end{center}
\end{table*}

Let us consider the SMO with Pm$\bar{3}$m symmetry and G-type AFM spin configuration, in which
the magnetic interactions are the SE couplings between two adjacent Mn ions at site $i$ and $j$.
The SE interactions can be further decomposed onto the contributions from intra-layer and inter-layer
as schematically shown in Fig.~\ref{fig:superlattices}(a),
which denote the SE couplings through virtual hopping processes mediated by equatorial
and the apical oxygen atoms on the oxygen octahedron respectively,
\begin{equation}
E_{\rm spin}=\sum\limits_{i,j} { {\mathcal J}_{i,j}^{\parallel}\vec{S}_i} \cdot \vec{S}_j
+\sum\limits_{i,j} { {\mathcal J}_{i,j}^{\perp}\vec{S}_i} \cdot \vec{S}_j .
\label{eq:exchange}
\end{equation}
In order to elucidate its electronic origin, we then apply the recently
developed extended KK model, in which the SE interactions can be expressed as\cite{Wang2014,HWang2016}
\begin{equation}
{\mathcal J}_{i,j}=\sum_{\alpha, \alpha '}J_{\alpha, \alpha '}^{{\rm AFM}}
- \sum_{\alpha, \beta}J_{\alpha , \beta}^{{\rm FM}}
\label{eq:hopping}
\end{equation}
The first term in Eq.~(\ref{eq:hopping}) describes the AFM-type coupling energy via a
virtual hopping process between the two half-filled $t_{2g}$ bands. The second
term depicts the competing FM-type coupling energy  originating from a hopping process
from the half-filled $t_{2g}$ bands to the empty $e_{g}$ bands.
Here, we adopt the convention to represent the AFM-type coupling and the
competing FM-type coupling energies by positive and negative signs respectively.
In our extended KK model, the coupling magnitude is proportional to the hopping integral $t^2$ and
electronic screening is considered by the constrained random phase approximation.
Both $t_{2g}$ and $e_{g}$ orbitals as well as their hopping integrals
are constructed based on MLWFs. The details of construction of our extended KK model can be
found in Ref.~\cite{Wang2014}.

In the unperturbed cubic SMO, the SE interactions are dominated by the AFM-type hopping
processes from the  half-filled $t_{2g}$ orbital on one Mn atom to another $t_{2g}$ orbital of
the neighboring Mn atoms. With the $t_{2g}$ states represented by the MLWFs, mixed oxygen
$2p$ character can be clearly identified in Fig.~\ref{fig:superlattices}(c)-(f).
This is consistent with the fact that the
SE interaction is mediated by the oxygen atom.
On the other hand, the FM-type hopping processes, with an opposite sign of coupling energy,
are all varnishing due to the orthogonal condition in the cubic symmetry.
In order to study the SPC effect, we compute the $J_{\alpha, \beta}$
as functions as various phonon modes amplitudes $u$ that have been
frozen into the cubic SMO perturbatively.
As shown in Fig.~\ref{fig:superlattices}, the quadratic dependence of the hopping integral
on the phonon amplitudes enable us to use the quadratic coefficient
$J^{''}_{\alpha, \beta}={\partial^2 J_{\alpha, \beta}}/{\partial u^2}$
to measure the spin-phonon coupling strengths for the individual hopping process.
The resulting $J^{''}_{\alpha, \beta}$ and total $J^{''}_{\rm T}$ are presented in Table~\ref{Tab:response}.

Consider $J^{''}_{\rm T}$ as the measure of the SPC strength, the negative values
of $J^{''}_{\rm T}$ indicate that the structural distortions
by all the four phonon modes suppress the AFM ordering energies, and favor the
stabilization of FM spin configurations. A close inspection further reals that
the magnitude of $J^{''}_{\rm T}$ thus the SPC strength decreases fast in order
of Slater, AFD, Axe, and the last mode. This feature is exactly consistent with the
the first-principles results based on the computed frequency shifts as shown in Table~\ref{Tab:response}.
However, our method can further unveils the electronic origins that are not accessible
in direct first-principles calculations as we discuss in the following.

Among all the phonon modes under investigation, the Slater mode has the largest SPC
strength. Such a large SPC effect originates from the rapidly decreased
SE interactions of both intra-layer coupling ($J^{''}_{\parallel}$) and inter-layer
coupling ($J^{''}_{\perp}$) as shown in Table~\ref{Tab:response}.
We first focus on the intra-layer electronic processes contributing to the SPC.
Under the Slater mode, an electric dipole is generated by the
relative displacements of the Mn cation and the octahedron in opposite directions
along the [001] axis as schematically shown in Fig.~\ref{fig:superlattices}(b).
In the above, the octahedron is moving approximately rigidly together with its six oxygen atoms.
As a result, the intra-layer Mn-O-Mn bonds are no longer straight lines as they were
in the cubic phase. Because of the above perturbed local chemical environment,
the $t_{2g}$ orbitals on two neighboring Mn atoms are distorted and tilted away
from each other as shown Fig.~\ref{fig:superlattices}(c).
Not surprisingly, the AFM-type SE coupling energies are decaying fast with the increased mode amplitude.
It can be also clearly seen by the relatively large magnitudes of
$J^{''}_{\parallel,(d_{xz},d_{xz})}$ and $J^{''}_{\parallel, (d_{yz},d_{yz})}$ in Table~\ref{Tab:response}.
Since the polar distortion is along the [001] direction, the intra-layer hopping processes
involving the $d_{xz}$ and  $d_{yz}$ orbitals are thus the most effectively coupled to the Slater mode.
The hopping process from $d_{xy} \rightarrow d_{xy}$
involving orbitals that spread out in the $xy$ plane is much less affected. Therefore, it is much
weakly coupled to to the Slater mode resulting in a relatively small
magnitude of  $J^{''}_{\parallel,(d_{xy}, d_{xy})}$. Interestingly,
the $e_g$ and $t_{2g}$ states are no longer orthogonal to each other by the broken symmetry under Slater mode.
As a result, the FM-type SE coupling is now allowed
and its coupling energies are increased rapidly with the mode amplitude as shown in Fig.~\ref{fig:superlattices}(c).
The above increased FM-type SE couplings, such as $J^{''}_{\parallel, (d_{xz},d_{x^2-y^2})}$, 
greatly contribute to the SPC effect.
In the next, we discuss the inter-layer electronic processes contributing to the SP.
The only non-varnishing contributions are due to the AFM-type hoping integrals of
$d_{xz} \rightarrow d_{xz}$ and $d_{yz} \rightarrow d_{yz}$ mediated by the apical oxygen atoms
on the octahedron. Under the Slater mode, the Mn atom is moving closer to one of the
apical oxygen atom while moving further away from the other apical oxygen atom.
As a result, the hybridized oxygen $2p$ character on the $d_{yz}$  or $d_{xz}$
is enhanced on one end, however, significantly suppressed on the other end.
Such a change in the local chemical environment reduces the effective overlap
of the $t_{2g}$ states and the SE coupling energies.

The AFD mode has the second largest SPC effect among the four phonon modes.
The AFD mode is a nonpolar structural distortion describing the oxygen octahedral rotation
around [001] axis. Under this mode, all the atomic displacements take place within the $xy$ plane.
Thus, it is expected that the SE interactions involving the $t_{2g}$ or $e_g$ electrons, whose main
orbitals are distributed within the $xy$ plane, will be most affected.
Indeed, a close inspection in Table~\ref{Tab:response} reveals that the SPC effect of this mode
is mainly contributed by $J^{''}_{\parallel, (d_{xy},d_{xy})}$.
Under this oxygen octahedron rotation mode, the $d_{xy}$ states on the two neighboring Mn atoms are
rotated away from each other, which reduces the hopping integral and the SE coupling energy as shown in
Fig.~\ref{fig:superlattices}(d). By the same token, the FM-type hopping process between $d_{xy}$ and the empty $e_g$ states
are no longer zero due to the symmetry breaking, which also enhance the FM-type
SE couplings and contribute to the overall SPC strength.

The Axe mode has a weaker SPC effect that that of the AFD mode. As schematically shown in Fig.~\ref{fig:superlattices}(b),
the Axe mode describe a polar distortion in which the electric dipole is generated by the relative displacements of the
apical and equatorial oxygen atoms in opposite directions along [001]. In the above, the two apical oxygen atoms
have a much large displacement than that of the equatorial oxygen atoms.
As a result, the SE coupling via the inter-layer Mn-O-Mn hopping processes are most perturbed by the Axe mode.
Indeed, the SPC effect of Axe mode are mainly contributed by the $J^{''}_{\perp, (d_{xz},d_{xz})}$  and
$J^{''}_{\perp, (d_{yz},d_{yz})}$ in Table~\ref{Tab:response}.
The above two terms describe the suppressed AFM-type SE coupling energies
when the Mn atom is moving closer to one of the apical oxygen atom but leaving away from
the other one in a similar way of how the inter-layer hopping processes contribute to the SPC of Slater mode as
we discuss previously.

Finally, we discuss the SPC effect by Last mode, which is the weakest among all the four modes.
This is due to the nature of the Last mode, in which an electric dipole is generated by the
displacements of the A-site atom moving in [001] direction and the octahedron with all six oxygen atoms
and the Mn atom together moving in [00$\bar 1$] direction as schematically shown in Fig.~\ref{fig:superlattices}(b).
Such a structural distortion barely modifies the local Mn-O bonding environment, which is
crucial for the SE coupling energy. As a result, both the $t_{2g}$ and the $e_g$ orbitals
are nearly intact due to the rigid displacement of the entire octahedron shown in Fig.~\ref{fig:superlattices}(f).
Not surprisingly, the SE coupling are not changed resulting in the observed
very weak SPC effect as shown in Table~\ref{Tab:response}.

\begin{table}[ht]
\begin{center}
\caption{The quadratic coefficients (meV/$\AA^2$)  of individual(total)
$J^{''}_{\parallel,(\alpha, \beta)}$ ($J^{''}_{\parallel}$) intra-layer
and $J^{''}_{\perp,(\alpha, \beta)}$ ($J^{''}_{\perp}$) inter-layer superexchange couplings
as functions of frozen mode amplitudes for the Slater mode in BiFeO$_3$.}
\begin{tabular}{c|cccccc|c}
  \hline
  \hline
       \scriptsize{AFM type}  &    & $d_{xy}$   & $d_{yz}$ & $d_{zx}$  & $d_{x^{2}-y^{2}}$ & $d_{z^{2}}$
& \multirow{5}{*}{{$J^{''}_{\parallel} =$ -5.7  }} \\
  \hline
 \multirow{5}{*}{$J^{''}_{\parallel,(\alpha, \beta)}$} &  $d_{xy}$                  & -4.0       &$\sim$0   & 0         & 0     & 0     \\
 & $d_{yz}$                  & $\sim$0    &-18.2     & 0         & 17.6 & 9.2  \\
 & $d_{zx}$                  & 0          & 0        &-18.2      & 17.6 & 9.2  \\
 & $d_{x^{2}-y^{2}}$        & 0          & 30.4    & 30.4     & -46.6 & -26.0 \\
 & $d_{z^{2}}$       & 0          & 15.9    & 15.9     & -26.0 & -12.9 \\
  \hline
 \multirow{5}{*}{$J^{''}_{\perp,(\alpha, \beta)}$}&  $d_{xy}$               & $\sim$0    & 0        & 0         & 0     & 0
 & \multirow{5}{*}{{$J^{''}_{\perp}$ = -64.3  }}   \\
 & $d_{yz}$               & 0          & -20.1    & 0         & 0     & 0    \\
 & $d_{zx}$               & 0          & 0        & -20.1     & 0     & 0     \\
 &$d_{x^{2}-y^{2}}$     & 0          & 0        & 0         &$\sim$0& 0     \\
 &$d_{z^{2}}$    & 0          & 0        & 0         & 0       & -24.1 \\
  \hline
\end{tabular}
\label{tab:BFO}
\end{center}
\end{table}

Based on the GKA theory, the metal-oxygen-metal angle has often been applied as a thumb of rule
to explain the SPC effect. Indeed, part of our analysis in the Slater
and AFD modes is consistent with the expectation from the above rule.
In fact, Slater mode and AFD mode are the two phonon
modes that directly change the Mn-O-Mn angles and they are also identified to
have the strongest SPC effect. Yet, the fact that slater mode has a much larger SPC
effect than that of AFD mode can not be satisfactorily explained by this phenomenological rule only,
since they adjust the Mn-O-Mn angle by a similar magnitude at the same mode amplitude\cite{HWang2016}.
Our current analysis provides a clearer electronic insight into the above discrepancy.
It shows that the Slater mode has a greater number of effectively coupled hopping processes
than AFD mode, which are crucially dependent on the symmetry of crystal field splitting.
In addition, the SPC in Slater mode benefits further from the inter-layer coupling.
In the above, the Mn-O-Mn remains a straight line, but the oxygen atom displaces
from the unperturbed symmetric position. This effect is not captured by the Mn-O-Mn angle argument.

The SPC effect in perovskite BFO appears even more elusive to the phenomenological rule
based on Fe-O-Fe angles. One would expect that the SPC of Slater mode should be comparable to that in
SMO, since the characteristic of the mode eigenvector is very similar to that in SMO.
On the contrary, first-principles calculations show a very weak SPC effect, in which the
frequency shift between AFM and FM spin configurations is only $\Delta \omega = 3 ~\rm{cm}^{-1}$.
This seemingly discrepancy can be easily explained by the analysis developed in this work.
In Table~\ref{tab:BFO}, we present the similar quantities of $J^{''}$ as a measure of
the SPC strengths from various electronic hopping processes in BFO.
As we have discussed earlier, the large SPC effect of Slater mode in SMO
benefits from both the rapidly suppressed AFM-type SE interaction between two half-filled
$d$ states and the same rapid increased FM-type SE interaction between one half-filled $d$ state
to the empty $d$ state. In sharp contrast to the electronic configuration of Mn$^{3+}$ in SMO
with half-filled $t_{2g}$ and empty $e_{2g}$ states, the Fe$^{3+}$ in BFO has both $t_{2g}$
and $e_{2g}$ states being half-filled. 
As a result, only AFM-type SE couplings in Eq.~\ref{eq:hopping} are allowed in BFO.
The FM-type hopping processes from $t_{2g}$ to $e_{2g}$ which contribute significantly to
the SPC effect of Slater mode in SMO now change their signs of energy as shown in Eq.~\ref{eq:hopping}
and become AFM-type. Instead of promoting SPC effect in SMO,
those SE coupling channels now largely suppress the SPC in BFO as shown by the
negative values of $J^{''}_{\alpha, \beta}$ from $t_{2g}$ to $e_{2g}$
hopping processes in Table~\ref{tab:BFO}. Therefore, the intra-layer coupling $J^{''}_{\parallel}$
is much smaller for Slater mode in BFO than in SMO.

In conclusion, we have studied the electronic origins of the spin phonon coupling
effect in transition metal perovskite by using the Wannier-based extended Kugel-Khomskii model.
It shows that the number of effectively coupled
electronic hopping processes is the key in the spin phonon coupling strength.
Those effectively coupled magnetic interactions are crucially
dependent on both the characteristic of the phonon mode and the $d$ orbitals in the
crystal splitting field. In perovskite, the phonon mode such as the Slater mode
which affects almost every metal-oxygen hybridization environment will maximize the SPC effect.
Following the same picture, it will not be difficult to understand that the ``Slater-like'' mode
($\Gamma_1^-$) is found to have strong SPC in double perovskite La$_2$NiMnO$_6$~\cite{HWang2016}.
The electronic configurations are important as well, which are
at variance in perovskite materials with different B-site cations.
The empty $d$ states of Mn atom in SMO play the decisive role in the fact that
its SPC effect is much stronger than that in BFO  with all half-filled $d$ states.
Following the above argument, the much larger
SPC effect in LaCrO$_3$ than that in LaFeO$_3$~\cite{Hong2012,HWang2016} reported in literature is not surprising.
In addition, a recent experiment in Bi$_2$FeCrO$_6$~\cite{Kamba2008} suggested a spin-phonon coupling
could be induced when the empty $d$ states are introduced by Cr to replace Fe, which
is also consistent with the conclusion in this work.

\section {Acknowledgments}
This work was supported as part of the Center
for the Computational Design of Functional Layered Materials,
an Energy Frontier Research Center funded by the U.S. Department
of Energy, Office of Science, Basic Energy Sciences under Award no.
DE-SC0012575 (X. W. designed the project and prepared the manuscript). 
L.H. was supported by the Chinese National Science Foundation
Grant number 11374275, and the  National Key Research and Development Program of China under Grants No. 2016YFB0201200 
(H. W. and L. H. constructed the extended Kugel-Khomskii model).
H. J. was supported by the National Natural Science Foundation (Project Nos. 21373017, 21321001)
and Ministry of Science and Technology  (2013CB933400) of China (H. W. and H. J. carried out the calculations within 
constrained random phase approximation).
This research used resources of the National Energy Research Scientific Computing Center (NERSC),
a DOE Office of Science User Facility supported by the Office of Science of the U.S.
Department of Energy under Contract No. DE-AC02-05CH11231.
X.W. thanks useful discussion with Craig Fennie.

$^*$ To whom correspondence should be addressed: xifanwu@temple.edu, and helx@ustc.edu.cn.

\end{document}